\documentclass[preprints,article,submit,pdftex,moreauthors]{Definitions/mdpi} 

\firstpage{1} 
\pubvolume{1}
\issuenum{1}
\articlenumber{0}
\pubyear{2025}
\copyrightyear{2025}
\datereceived{ } 
\daterevised{ } 
\dateaccepted{ } 
\datepublished{ } 
\hreflink{https://doi.org/} 

\usepackage{siunitx}
\DeclareSIUnit{\sample}{Sa}
\DeclareSIUnit{\electroncharge}{e}
\usepackage[siunitx]{circuitikz}
\Title{Extraction of Electron and Hole Drift Velocities in thin 4H-SiC PIN Detectors using High-Frequency Readout Electronics}

\TitleCitation{Extraction of Electron and Hole Drift Velocities in thin 4H-SiC PIN Detectors using High-Frequency Readout Electronics}

\Author{Andreas Gsponer$^{1,2}$*\orcidA{}, Sebastian Onder$^{1,2}$\orcidB{}, Stefan Gundacker$^{1}$\orcidC{}, Jürgen Burin$^{1}$\orcidD{}, Matthias Knopf$^{2}$\orcidE{}, Daniel Radmanovac$^{1,2}$\orcidF{}, Simon Waid$^{1}$\orcidG{} and Thomas Bergauer$^{1}$\orcidH{}}

\AuthorNames{Andreas Gsponer, Sebastian Onder, Stefan Gundacker, Jürgen Burin, Matthias Knopf, Daniel Radmanovac, Simon Waid and Thomas Bergauer}

\address{%
$^{1}$ \quad Marietta Blau Institute for Particle Physics, Austrian Academy of Sciences,  Dominikanerbastei 16, 1010 Vienna, Austria\\
$^{2}$ \quad Institute of Atomic and Subatomic Physics, TU Wien, Stadionallee 2, 1020 Vienna, Austria}

\corres{Correspondence: andreas.gsponer@oeaw.ac.at}

\abstract{
Silicon carbide (SiC) has been widely adopted in the semiconductor industry, particularly in power electronics, because of its high temperature stability, high breakdown field, and fast switching speeds.
Its wide bandgap makes it an interesting candidate for radiation-hard particle detectors in high-energy physics and medical applications.
Furthermore, the high electron and hole drift velocities in 4H-SiC enable devices suitable for ultra-fast particle detection and timing applications.
However, currently, the front-end readout electronics used for 4H-SiC detectors constitute a bottleneck in investigations of the charge carrier drift.
To address these limitations, a high-frequency readout board with an intrinsic bandwidth of \SI{10}{\giga\hertz} was developed.
With this readout, the transient current signals of a 4H-SiC diode with a diameter of \SI{141}{\micro\meter} and a thickness of \SI{50}{\micro\meter} upon UV-laser, alpha particle, and high-energy proton beam excitation were recorded.
In all three cases, the electron and hole drift can clearly be separated, which enables the extraction of the charge carrier drift velocities as a function of the electric field.
These velocities, for the first time directly measured, provide a valuable comparison to Monte-Carlo simulated literature values and constitute an essential input for TCAD simulations.
Finally, a complete simulation environment combining TCAD, the Allpix$^2$ framework, and SPICE simulations is presented, in good agreement with the measured data. 
}

\keyword{Silicon Carbide (SiC); high-frequency; drift velocity; transient-current-technique} 

\begin{document}

\section{Introduction}
Silicon carbide (SiC) is an attractive candidate for future detectors in particle physics experiments due to its wide bandgap and, therefore, extremely low dark currents, even after irradiation and at room temperature \cite{Gaggl2022, Gsponer_2023}.
The 4H-SiC polytype (with a bandgap of \SI{3.26}{\electronvolt}) has been investigated as a radiation detector material for over 60 years \cite{Babcock1965,nava2008, denapoli2022} and has recently received renewed interest in the community due to manufacturing advances in the power electronics industry.
Since the 1970s, the charge carrier drift properties inside semiconductors have been studied by injecting charge carriers and evaluating the transient currents ~\cite{jungclaussen_analyse_1967, zulliger_electric_1969}, for example, to extract drift velocities as a function of the electric field and temperature~\cite{canali_drift_1971}.
In the last 30 years, this technique has also been applied very successfully to investigate solid-state detectors in high-energy physics, characterizing radiation damage-induced changes of the electric field in detectors~\cite{eremin_development_1996, li_direct_1997, Kramberger2010}, and the term transient-current-technique (TCT) has been coined.
The material parameters of 4H-SiC are not as well known as for silicon, and significant uncertainties still exist, for example, in the high-field mobility models (especially for holes) \cite{burin2025}.
One shortcoming of 4H-SiC is currently the limited epitaxial thicknesses that can be grown and depleted, typically \SI{100}{\micro\meter} or less \cite{denapoli2022}.
Combined with an electron saturation drift velocity that is approximately twice as high as for silicon, this results in extremely fast signals.
Therefore, in order to characterize the transient current in 4H-SiC, much faster readout electronics than currently used to characterize silicon detectors are required.

This paper presents a novel readout scheme using a monolithic microwave integrated circuit (MMIC) based low-noise amplifier adapted to read out \SI{50}{\micro\meter} thick 4H-SiC samples with low capacitance.
The maximum readout bandwidth of \SI{10}{\giga\hertz} makes it possible to resolve the transient current induced by charge carrier drift inside the detector for the first time experimentally.
Charge carriers were injected with different ionization profiles, using a \SI{370}{\nano\meter} pulsed UV-laser, alpha particles, and proton beams at a medical synchrotron (MedAustron), and the transient currents are compared.

\section{Materials and Methods}
In solid-state detector applications, the quest for higher readout bandwidths has been driven mainly by timing applications, such as silicon low-gain avalanche diodes (LGADs)~\cite{lgad1,lgad2} or silicon photomultipliers (SiPMs) \cite{Gundacker_2019, KRAKE2022167032}.
However, timing applications do not necessarily call for the highest possible bandwidth, but instead limit the bandwidth to the frequency content of the detector signal in order to achieve the best slope-to-noise ratio and time resolution.
If, however, the signal-to-noise ratio (SNR) is sufficiently high, for example, when a large amount of charge is deposited (greater than that of a minimum ionizing particle) or when waveform averaging is used, the readout bandwidth can be further increased and the transient charge carrier drift can be resolved in detail.

Typical silicon-based timing detectors in high-energy physics applications have capacitances of \num{1}-\SI{2}{\pico\farad} and a readout bandwidth below \SI{2}{\giga\hertz}, in accordance with the length of the detector signal~\cite{lgad2}.
To increase the bandwidth (and decrease signal rise-times), either the detector capacitance or the input impedance of the amplifier has to be reduced.
For the latter, approaches using shunt-resistors exist, reducing the input impedance to \SI{10}{\ohm}~\cite{Schumm_2024}.
However, resistances smaller than this are not practically feasible at \si{\giga\hertz} frequencies, and most of the signal is lost through the shunt.
Another way to reduce the input impedance is to use active feedback.
Nevertheless, the bandwidths that can be achieved in this approach are limited due to signal transfer delays.
The detector capacitance is a result of the sensor geometry and, in general, for \SI{50}{\micro\meter} thick semiconductor detectors, the electrically active area should be below \SI{0.5}{\milli\meter^2} to achieve capacitances of below \SI{1}{\pico\farad}, which corresponds to signal bandwidths above \SI{3}{\giga\hertz} for an input impedance of \SI{50}{\ohm}.

The following sections describe the 4H-SiC sensor used in the measurements, the high-frequency readout design, and a Monte-Carlo simulation framework based on technology computer-aided design (TCAD), Allpix$^2$, and SPICE simulations to model the measured transient signals.
\begin{figure}[htp]
    \isPreprints{\centering}{}
    \includegraphics[width=0.7\linewidth]{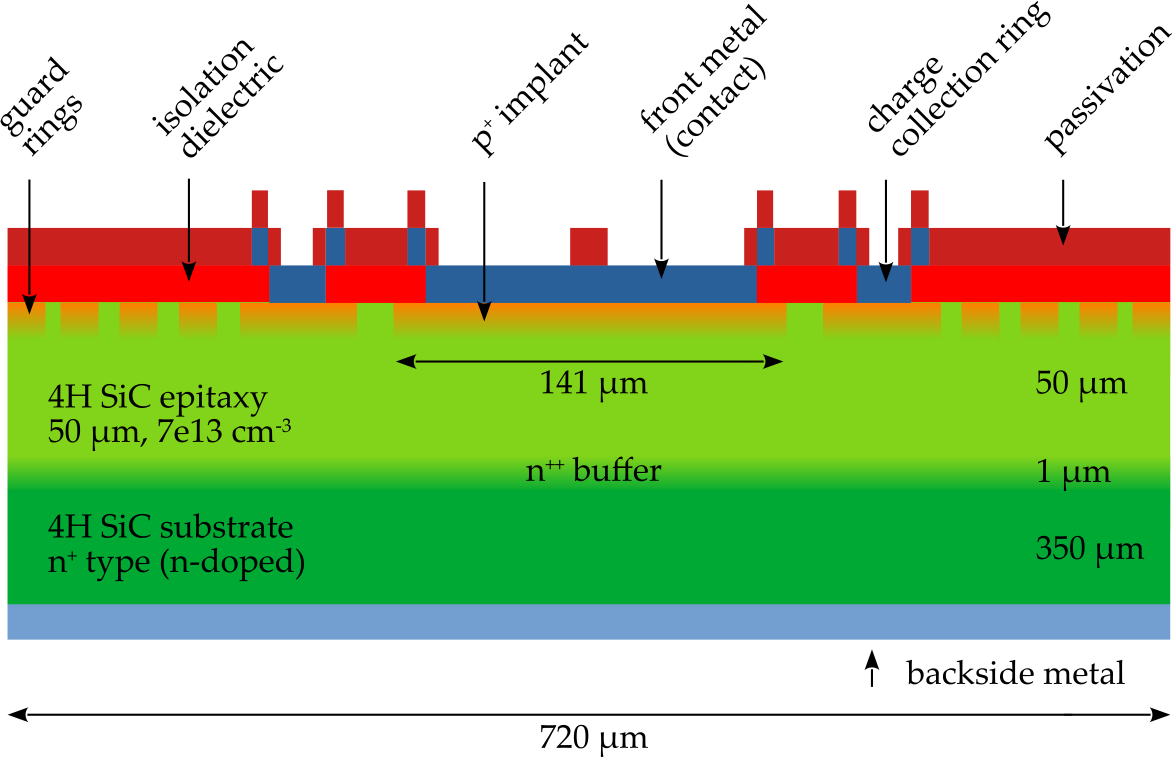}
    \caption{Diode cross-section showing the p-n junction, low-doped epitaxial layer, highly-doped substrate, pad (contact), and guard rings.}
    \label{fig:materials:cross_section}
\end{figure}

\subsection{Silicon Carbide Sensors}
\begin{figure}[htp]
    \isPreprints{\centering}{}
    \begin{minipage}[b]{0.35\textwidth}
        \includegraphics[height=.18\textheight]{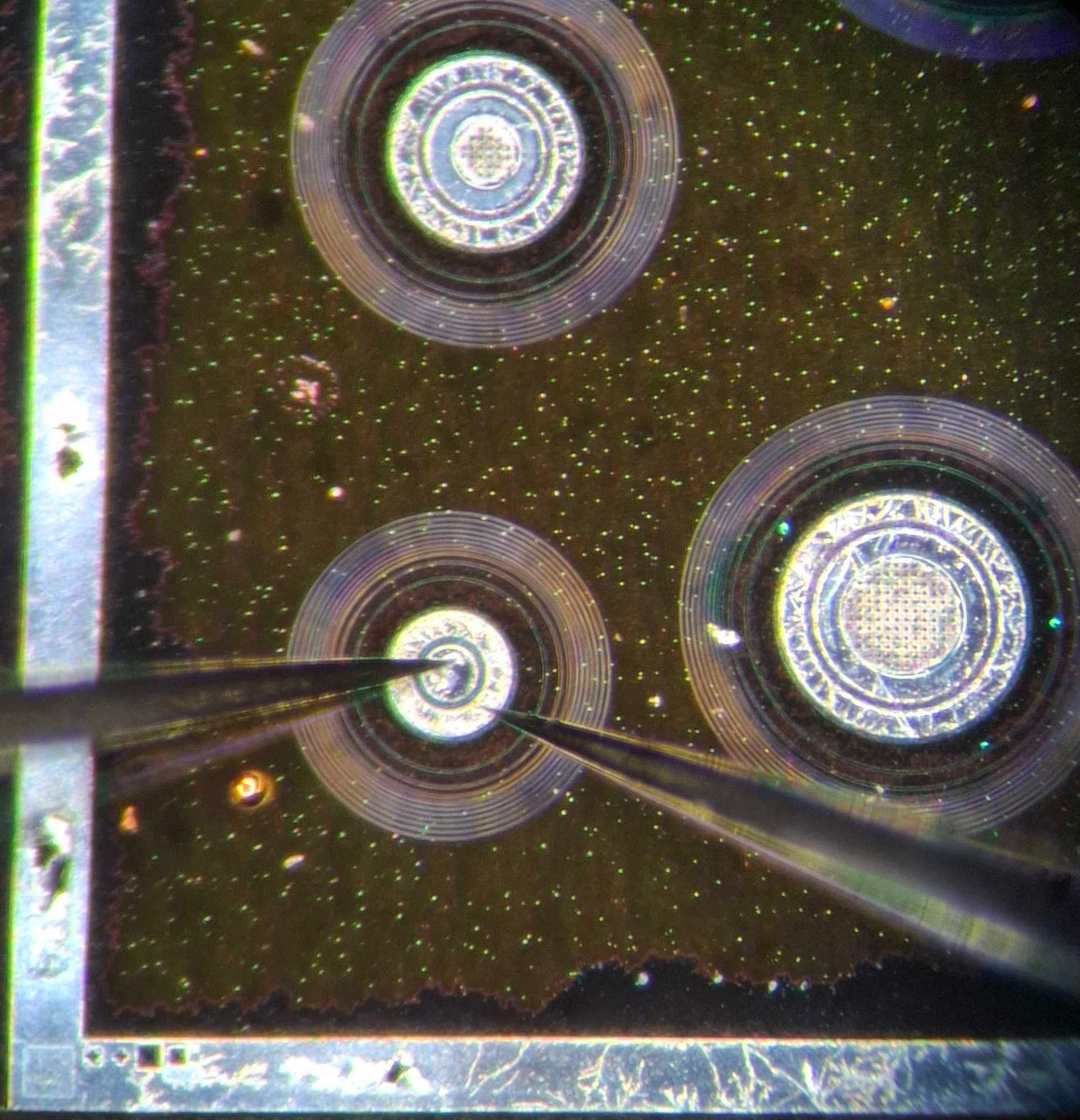}
        \quad \\[0.1em]
        \centering (a)
    \end{minipage}
    \hspace{{0.07\textwidth}}
    \begin{minipage}[b]{0.54\textwidth}
        \includegraphics[height=.18\textheight]{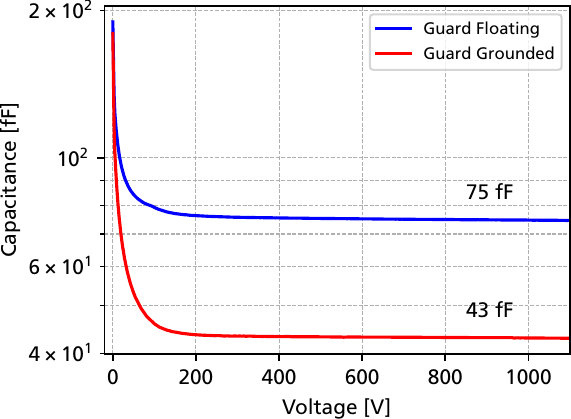}
        \quad \\[0.1em]
        \centering (b)
    \end{minipage}
    \caption{\textbf{(a)}: Photograph of the 4H-SiC diode undergoing electrical characterization. \textbf{(b)}: Measured capacitance as a function of the bias voltage when the guard ring is floating or grounded.}
    \label{fig:materials:ivcv}
\end{figure}

The silicon carbide sensors used for this study were manufactured by IMB-CNM, Barcelona, as a part of the RD50-SiC-LGAD common project.
They were produced on a 6-inch wafer with an epitaxial layer of \SI{50}{\micro\meter} thickness and an n-type doping of approximately \SI{7e13}{\per\centi\meter^{-3}} on a \SI{350}{\micro\meter} substrate.
An image of the sensor cross-section is shown in Figure~\ref{fig:materials:cross_section}.
Diodes with a small area (\SI{0.016}{\milli\meter^2}) were chosen to achieve a very low capacitance and to maximize the achievable bandwidth.
Figure~\ref{fig:materials:ivcv} shows an image of the diode undergoing electrical characterization and the measured capacitance as a function of the bias voltage.

Full depletion of the device is reached after around \SI{100}{\volt}, with a capacitance of \SI{75}{\femto\farad} (for a floating guard ring).
The bias current stays well below \SI{100}{\pico\ampere} for all bias voltages and can be attributed to surface currents, since thermal generation of charge carriers in the epitaxial 4H-SiC is negligible at room temperature.
The maximum voltage in the experimental measurements was limited to \SI{1.1}{\kilo\volt}.
This limitation is not due to a breakdown in the sensor itself (as 4H-SiC can withstand fields up to \SI{3}{\mega\volt\per\centi\meter}), but a result of electric breakdown in the air between the edge of the diode and the ground plane of the readout electronics.
Different techniques, such as ultra-high vacuum or encapsulation, could be used in the future to mitigate these problems.
Additionally, thinner detectors could be used to access higher field regions using the same bias voltage ranges.

\subsection{High-Frequency Readout Electronics}
\begin{figure}[htp]
    \isPreprints{\centering}{} 
    \begin{minipage}[t]{0.41\textwidth}
        \includegraphics[height=.22\textheight]{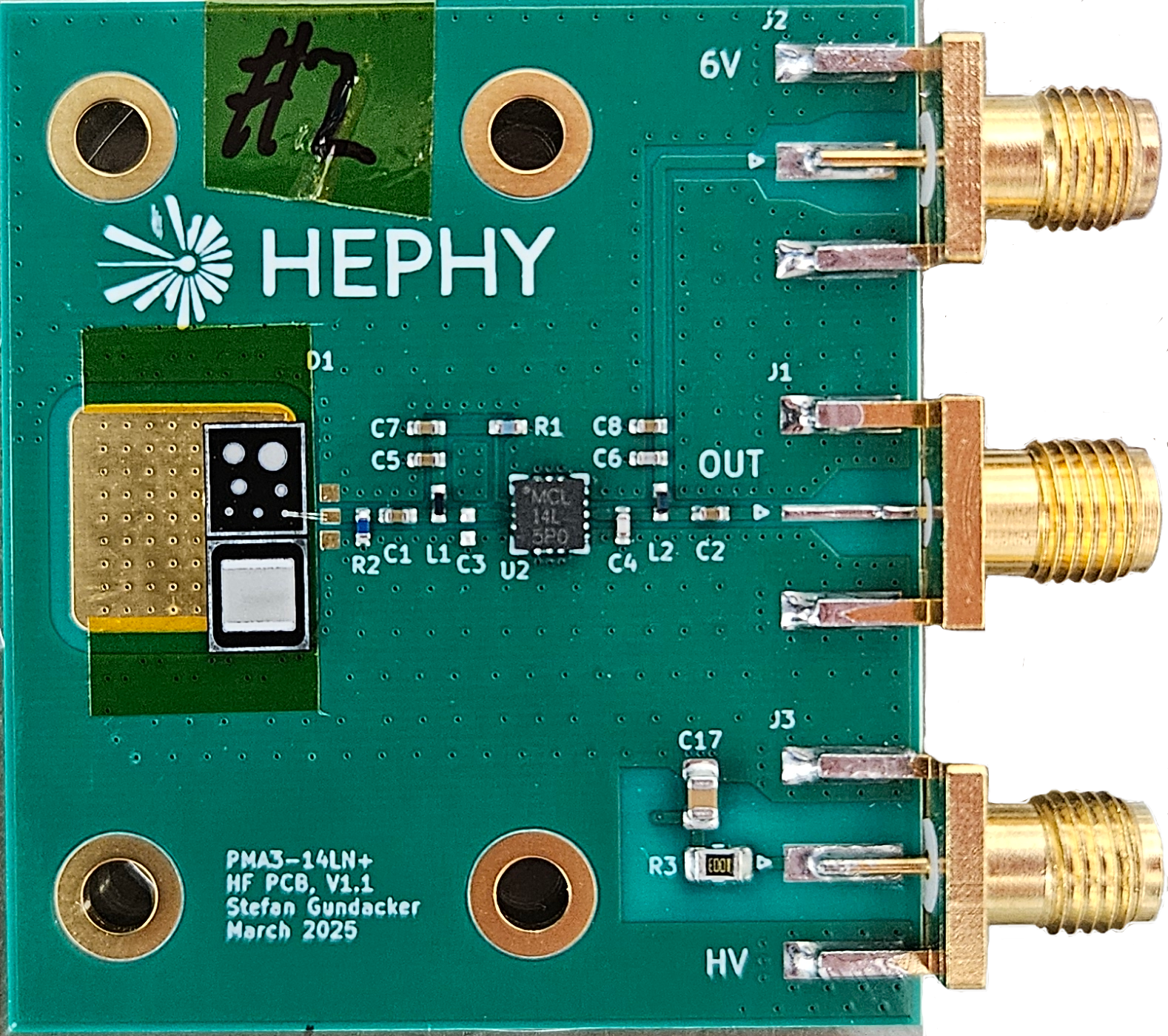}
        \quad \\[0.1em]
        \centering (a)
    \end{minipage}
    \hspace{0.02\textwidth}
    \begin{minipage}[t]{0.53\textwidth}
        \includegraphics[height=.22\textheight]{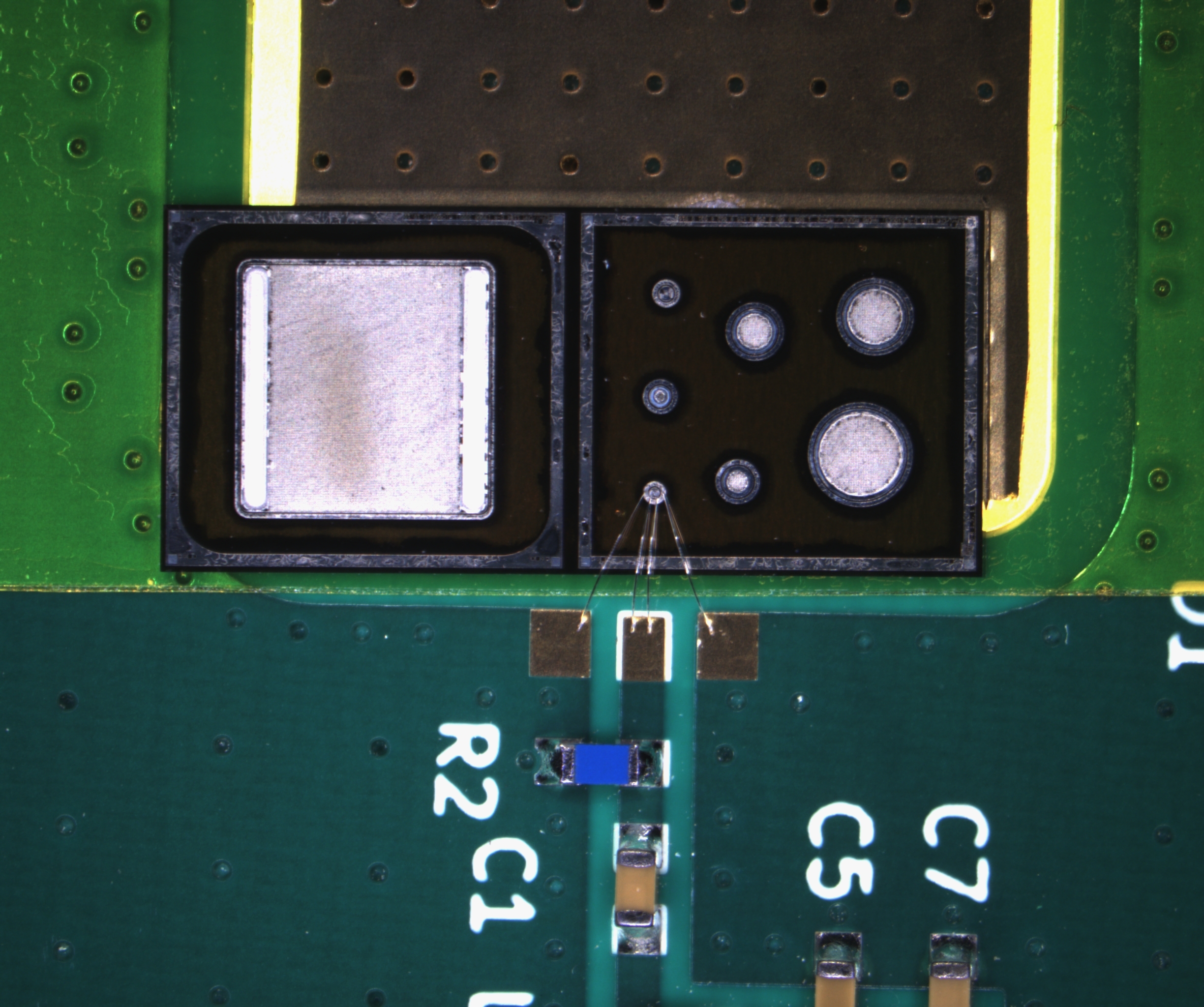}
        \quad \\[0.1em]
        \centering (b)
    \end{minipage}
    \caption{\textbf{(a)} Readout PCB with mounted detector. \textbf{(b)}: Microscope image of the detector on the PCB, wirebonded to a transmission line. In the measurements, the diode's guard ring was not connected.}
    \label{fig:materials:microscope}
\end{figure}

The core of the readout is an MMIC-based low noise amplifier (PMA3-14LN+) from Mini-Circuits, with a bandwidth of \SI{50}{\mega\hertz}-\SI{10}{\giga\hertz}, a flat gain of \SI[separate-uncertainty=true]{22.6(7)}{\decibel} and a low noise figure of around \SI{1.1}{\decibel} in the frequency range of interest \cite{minicircuitdatasheet}.
A printed circuit board (PCB) based on Rogers 4350B substrate was designed to accommodate the MMIC, depicted in Figure~\ref{fig:materials:microscope}.
The detector capacitance and inductance of the wire bonds form a low-pass filter defined by the amplifier's input impedance, limiting the bandwidth that can be achieved.
Furthermore, the inductance of the wire bond (typically \SI{1}{\nano\henry} per mm wire length) can create a damped oscillator with the detector capacitance.
For the detector used in the measurements, this resonance frequency is estimated to be around \SI{9}{\giga\hertz} and can negatively influence the signal transients. These effects will be investigated in more detail with dedicated SPICE simulations in Section~\ref{sec:qucs}.

The detector is wire bonded to a transmission line, where a \SI{10}{\nano\farad} capacitor and a \SI{100}{\kilo\ohm} resistor form a bias-tee.
The sensor die itself is glued to a high-voltage plane, which is connected to eleven 0805-size \SI{2.2}{\nano\farad}/\SI{1}{\kilo\volt} SMD capacitors to ground on the back side of the PCB, providing a signal return path as short as possible.
Multiple capacitors are used in parallel to minimize the effects of parasitic inductance.
All measurements were performed with a Rohde\&Schwarz RTP164 oscilloscope at a sample rate of \SI[per-mode=symbol]{40}{\giga\sample\per\second} and an analog bandwidth of \SI{16}{\giga\hertz}.
The measured voltage pulses $V(t)$ were converted to current pulses using $I(t)={V(t)}/{(A \cdot \SI{50}{\ohm})}$, assuming an amplification of $A=\num{13.5}$~(\SI{22.6}{\decibel}) and a \SI{50}{\ohm} input impedance of the MMIC.

\subsubsection{Feature Extraction}
The transient current induced by the charge carrier drift between two electrodes can be defined by the Shockley-Ramo theorem~\cite{Shockley1938, Ramo1939}
\begin{equation}
    I = q \cdot \vec{v} \cdot \vec{E_w}\,,
\end{equation}
where $q$ is the charge, $\vec{v}$ the drift velocity, and $\vec{E_w}$ the gradient of the \emph{weighting potential}.
For selective electron/hole injection at the top/bottom of the device, the resulting transient current will be a rectangular pulse, which can be directly used to measure the drift time (known as the "time-of-flight" technique)~\cite{Martini1972}.
However, for very thin detectors, selectively injecting electrons or holes using ionizing radiation is challenging.
For example, \SI{5.5}{\mega\electronvolt} alpha particles already penetrate more than \SI{15}{\micro\meter} into 4H-SiC~\cite{gsponer_measurement_2024}.

\begin{figure}[htp]
    \isPreprints{\centering}{}
    \includegraphics[width=0.625\linewidth]{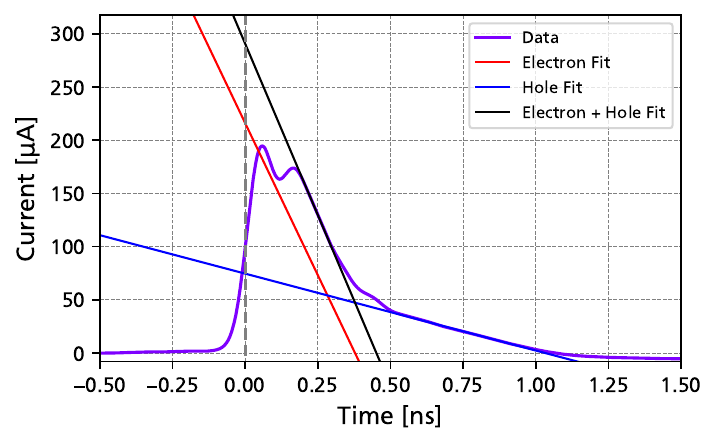}
    \caption{Measured triangle-like transient current induced by electron and hole drift in a pulsed UV-laser measurement together with linear fits used to extract charge carrier drift velocities in 4H-SiC.}
    \label{fig:materials:extraction}
\end{figure}

Instead, a uniform charge deposition along the depth of the detector can be used.
For a constant electric field (i.e., constant drift velocity), the transient current will then be of triangular shape for the electrons and holes, respectively~\cite{Kramberger2015}.
The electron and hole drift times ($t_e$ and $t_h$) can thus be extracted by linear fits of the two falling slopes and their zero crossings, as depicted in Figure~\ref{fig:materials:extraction}.
For bias voltages $V$ well above the full depletion voltage (where a constant electric field $E=V/d$ can be assumed), the charge carrier drift velocities can be calculated according to $v_{e,h}(E)=d/t_{e,h}(E)$, with $d=\SI{50}{\micro\meter}$ being the epitaxial thickness of the detector.
The electron and hole mobilities are then finally obtained by $\mu_{e,h}(E)=v_{e,h}(E)/E$.

\subsection{Simulation Framework}

In order to verify the results and compare the measured transient currents with theoretical models, electric field simulations in Sentaurus TCAD, Monte-Carlo simulations in Allpix$^2$, and SPICE simulations using QUCS-S have been performed.

\subsubsection{TCAD Simulations}
\begin{figure}[htp]
    \isPreprints{\centering}{} 
    \begin{minipage}[b]{0.34\textwidth}
        \includegraphics[height=.19\textheight]{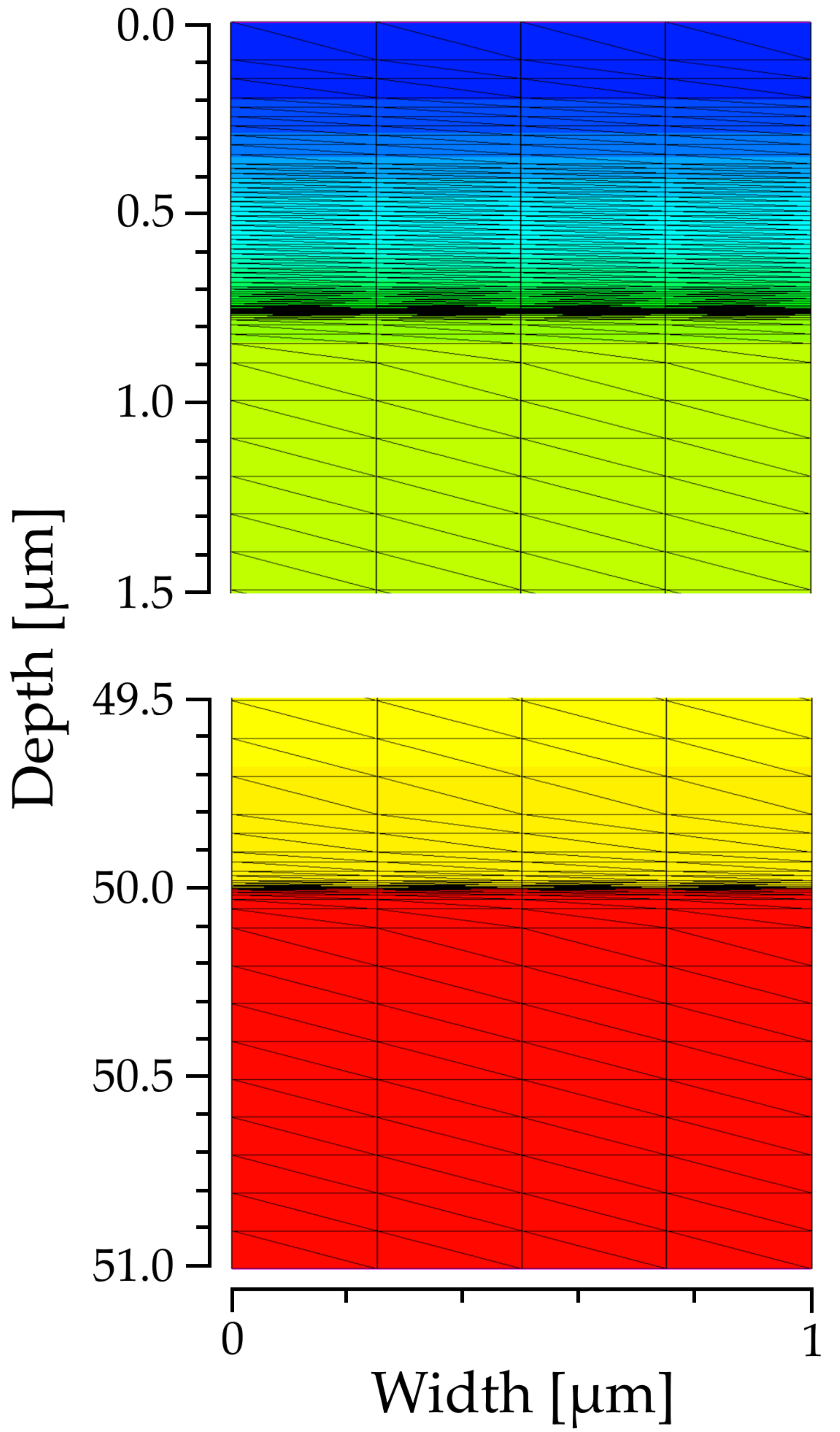}
        \quad \\[0.1em]
        \centering (a)
    \end{minipage}
    \hspace{{0.02\textwidth}}
    \begin{minipage}[b]{0.6\textwidth}
        \includegraphics[height=.19\textheight]{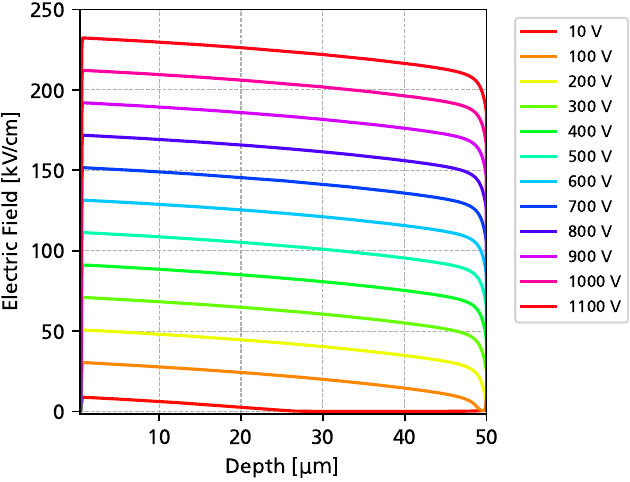}
        \quad \\[0.1em]
        \centering (b)
    \end{minipage}
    \caption{\textbf{(a)}: TCAD simulation geometry, with the color representing the effective doping concentration. \textbf{(b)}: Resulting simulated electric fields as a function of the bias voltage.}
    \label{fig:tcadfield}
\end{figure}

In a first step, the 4H-SiC diode was simulated using Synopsis Sentaurus TCAD in order to extract electrostatic potentials and electric field profiles intended for import into the Allpix$^2$ simulations.
The electric fields were simulated using a quasi 1-D geometry using the device depth and a width of \SI{1}{\micro \meter}, as depicted in Figure~\ref{fig:tcadfield}(a).
By omitting edge termination structures, which have a negligible effect on the electric field in the center of the diode, computation time could be significantly reduced.
The p$^{++}$ implant was specified using measurements performed by the manufacturer, and the doping concentration of the intrinsic epitaxial layer was extracted from the C-V measurements (see Figure~\ref{fig:materials:ivcv}(b)).
At the bottom of the geometry, a $\SI{1}{\micro \meter}$ thick buffer layer with a doping concentration of $\SI{1e18}{\centi \meter^{-3}}$ was added in the form of a box profile.
The simulation setup, including material parameters, material anisotropy, and solving criteria, matches previously published simulation work done by the authors \cite{URL_SiC_Sim, Onder2025}.
From the simulations, the electrostatic potentials and corresponding electric fields at bias voltages from $\SI{0}{\volt}$ up to $\SI{1100}{\volt}$ were calculated.
Selected electric field distributions along a cut across the detector thickness are shown in Figure~\ref{fig:tcadfield}, showing non-linear electric fields due to the non-constant epitaxial doping profiles.

\subsubsection{Allpix$^2$ Simulations}
In order to simulate the response of the detector to ionizing radiation (and to take into account the stochastic energy loss), the Allpix$^2$ framework~\cite{Allpix} was used.
The electric fields and the electrostatic potentials from the TCAD simulations were exported, converted, and subsequently imported into Allpix$^2$.
The weighting field was calculated by subtracting the electrostatic potential from two simulations with a \SI{1}{\volt} difference.

For the simulation of alpha particles hitting the sensor, the \textit{DepositionGeant4} module was used, while the UV-laser signals were simulated using the \textit{DepositionLaser} module.
To simulate the laser with a wavelength of \SI{370}{\nano \meter}, an absorption coefficient of \SI{42.25}{cm^{-1}} \cite{Sridhara1998} was used, with a cylindrical beam geometry and a pulse duration of \SI{25}{\pico \second} (approximately \SI{58}{\pico \second} FWHM).
\num{100000} photons were simulated, and the resulting signal was scaled to match the magnitude of the measurements.

Charge carrier propagation was performed using the \textit{TransientPropagation} module, using \SI{0.5}{\pico \second} time steps and charge carriers propagated in groups of 100 per step.
Finally, the charge carrier transport parameters of 4H-SiC were implemented in Allpix$^2$ using a custom mobility model corresponding to the Caughey-Thomas formula:
\isPreprints{}{
\begin{adjustwidth}{-\extralength}{0cm}
} 
\begin{equation}
    \mu(E) = \frac{\mu_\mathrm{low}}{\left(1 + \left(\frac{\mu_\mathrm{low}E}{v_\mathrm{sat}}\right)^{\beta}\right)^{1/\beta}}\,,
    \label{eq:mobility}
\end{equation}
\isPreprints{}{
\end{adjustwidth}
} 
with a fixed low field mobility $\mu_\mathrm{low}$ of \SI{950}{\centi \meter^2 \volt^{-1} \second^{-1}} for electrons and \SI{115}{\centi \meter^2 \volt^{-1} \second^{-1}} for holes~\cite{Ishikawa2023, Ishikawa2024}.
The saturation velocity $v_\mathrm{sat}$ and the empirical coefficient $\beta$ are based on the measured parameters (see Section~\ref{sec:vsat_results}).
4H-SiC features an anisotropy (parallel/perpendicular to the $c$-axis) in many of its material parameters, including the mobility~\cite{burin2025}.
As the epitaxial layer of the detector is grown \SI{4}{\degree} off-axis (to ensure polytype control), and the drift of the charge carriers is primarily in one axis (along the detector thickness), the mobility parallel to the $c$-axis was used.

\subsubsection{QUCS simulations}
\label{sec:qucs}
\begin{figure}[htp]
    \isPreprints{\centering}{} 
    \includegraphics[width=.8\textwidth]{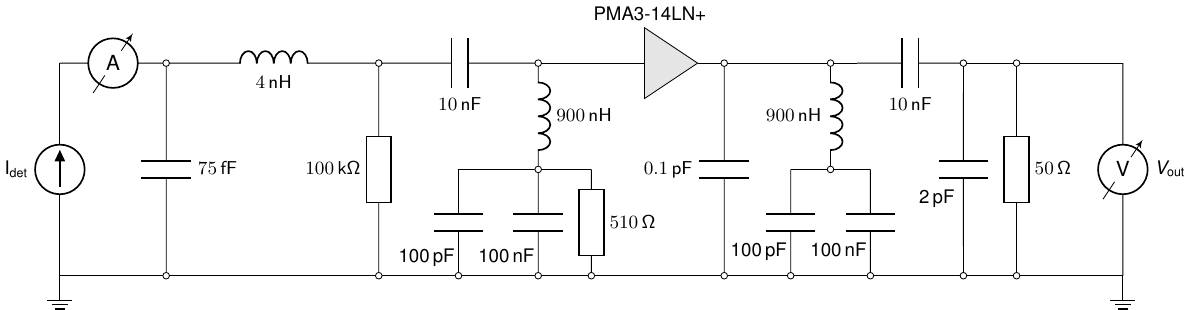}
    \caption{Simulation schematics and input parameters.}
    \label{fig:qucs}
\end{figure}

QUCS (quite universal circuit simulator) \cite{qucs_link} is an open-source SPICE (simulation program with integrated circuit emphasis) software.
In QUCS, the circuit schematic (shown in Figure~\ref{fig:qucs}) was included, and the wire bond was modeled as an inductance of \SI{4}{\nano\henry}.
The detector capacitance was specified to be \SI{75}{\femto\farad}, according to the C-V measurements in Figure~\ref{fig:materials:ivcv}.
The attenuation of the SMA cable connecting the readout electronics to the oscilloscope was phenomenologically modeled by including a \SI{2}{\pico\farad} capacitance parallel to the \SI{50}{\ohm} input impedance of the oscilloscope.
The input current $\mathrm{I_{det}}$, provided by the TCAD and Allpix$^2$ simulations described in the previous chapters, was included in QUCS via a file-based current source.
\section{Results}
\label{sec:measurements}
\subsection{Pulsed UV-Laser}
\label{sec:uvtct}
A PILAS PIL1-037-40FC picosecond pulsed laser with a wavelength of \SI{370}{\nano\meter} was used to inject charges inside the metallization opening in between the pad and the guard ring (compare with Figure~\ref{fig:materials:cross_section}), with a 1-$\sigma$ spot size of around \SI{28}{\micro\meter}.
The attenuation length of 4H-SiC for this wavelength is around \SI{240}{\micro\meter}~\cite{Sridhara1998}, which implies that photons are absorbed (and electron hole pairs created) almost uniformly along the thickness of the epitaxial layer.
The FWHM of the laser pulse was measured by the manufacturer to be \SI{58}{\pico\second}.
Additionally, the laser provides a very low jitter trigger output (verified to be less than \SI{4}{\pico\second} relative to the laser pulse), which has been used to trigger the oscilloscope and average \num{1000} waveforms per acquisition, reducing noise.
By changing the trigger threshold, the acquisition can be shifted in time, allowing the real-time oscilloscope to be used as a synthetic sampling oscilloscope with a sample rate in excess of \SI[per-mode=symbol]{40}{\giga\sample\per\second}.
\begin{figure}[htp]
    \isPreprints{\centering}{} 
    \includegraphics[width=.625\textwidth]{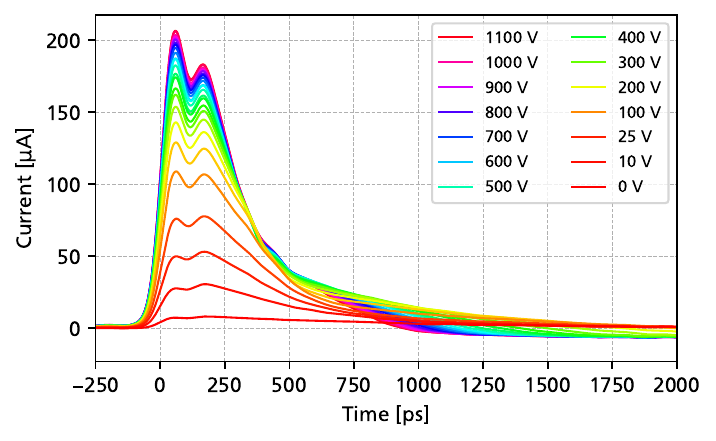}
    \caption{UV-TCT transient current as a function of the applied reverse bias voltage.}
    \label{fig:results:tct}
\end{figure}
Figure~\ref{fig:results:tct} shows the measured UV-TCT transient current waveforms as a function of the bias voltage.
At the maximum of the waveform, a "double-peak" can be observed, which can be attributed to a resonance of the wire bond inductance (discussed in further detail in Section~\ref{sec:qucs}).
For the data shown, a synthetic sample rate of \SI[per-mode=symbol]{400}{\giga\sample\per\second} has been used.
At the highest bias voltage, the total signal duration is shorter than \SI{1}{\nano\second}, with a FWHM of around \SI{300}{\pico\second}.
A clear separation of electron drift (first steep falling edge) and hole drift (slower signal tail) can be seen.
The amplitude of the signals increases even after full depletion is reached (around \SI{100}{\volt}), as the charge carrier drift velocities continue to grow with higher bias voltages.

\subsection{Proton Beams at MedAustron}
The MedAustron ion therapy center offers proton beams in the energy range between \SI{62.4}{\mega\electronvolt} and \SI{800}{\mega\electronvolt}.
In order to obtain the highest charge deposition in the detector, the lowest proton energy was chosen, corresponding to an energy loss of around \num{5} minimum ionizing particles (MIPs).
For the measurements, an additional \SI{22}{\decibel} ZX60-14LN-S+ low-noise amplifier from Mini-Circuits was used as a second stage to boost the signals above the noise floor of the oscilloscope.
Waveforms were recorded by splitting the signal into two channels, one channel recording the signal at the full analog bandwidth of \SI{16}{\giga\hertz} and the other serving as a trigger using a \SI{500}{\mega\hertz} low-pass filter.
For each bias voltage between \SI{100}{\volt} and \SI{1.1}{\kilo\volt}, \num{10000} acquisitions were obtained.

\begin{figure}[htp]
    \isPreprints{\centering}{} 
    \begin{minipage}[b]{0.3\textwidth}
        \includegraphics[width=\textwidth]{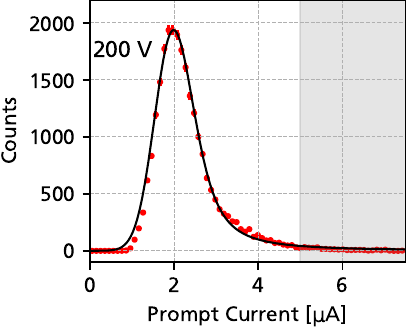}
    \end{minipage}
    \hspace{{0.01\textwidth}}
    \begin{minipage}[b]{0.3\textwidth}
        \includegraphics[width=\textwidth]{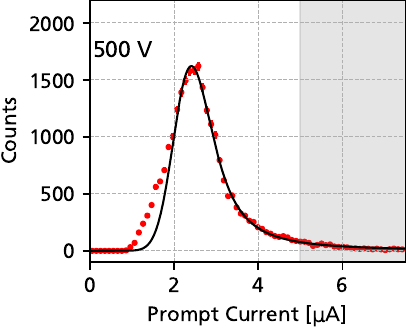}
    \end{minipage}
    \hspace{{0.01\textwidth}}
    \begin{minipage}[b]{0.3\textwidth}
        \includegraphics[width=\textwidth]{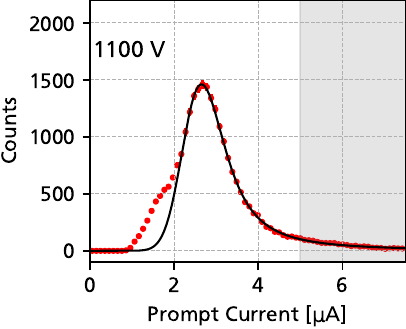}
    \end{minipage}
    \caption{Prompt current (proportional to the deposited energy) for \SI{62.4}{\mega\electronvolt} protons for different detector bias voltages. The distributions have been fitted using a Landau distribution, and the region used for waveform-averaging ($>\SI{5}{\micro\ampere}$) is indicated using gray shading.}
    \label{fig:results:proton:landau}
\end{figure}

Figure~\ref{fig:results:proton:landau} shows a histogram of the transient current waveform amplitude, proportional to the sum of the electron and hole drift velocity.
Due to stochastic fluctuations in the interaction of ionizing radiation, the energy loss of the protons follows a \emph{Landau} distribution, with a most probable value of around \SI{120}{\kilo\electronvolt} (or \SI{15}{\kilo\electroncharge}~\cite{gsponer_measurement_2024}) in \SI{50}{\micro\meter} 4H-SiC.
At a bias voltage of \SI{1}{\kilo\volt}, an SNR of around \num{5} is achieved.

The shape of the transient current waveforms is strongly affected by fluctuations and non-uniformities in the energy loss of protons, and in order to compare the observed charge carrier drift to UV-TCT measurements, waveform averaging is required.
However, due to the poor SNR, the rising edge of the signals cannot be aligned sufficiently accurately to prevent a "smearing-out" of the signals.
To overcome this limitation, waveform averaging was performed using only a subset of waveforms from the high SNR tail of the energy deposition distribution, as indicated by the gray shaded area (SNR > 10) in Figure~\ref{fig:results:proton:landau}.
Depending on the bias voltage, this resulted in about \num{500} waveforms available for averaging.
This "cherry-picking" detaches the information about the amplitude of the waveforms, which was recovered by scaling the averaged waveforms by the most-probable prompt current obtained from histograms.
\begin{figure}[htp]
    \isPreprints{\centering}{} 
    \includegraphics[width=.625\textwidth]{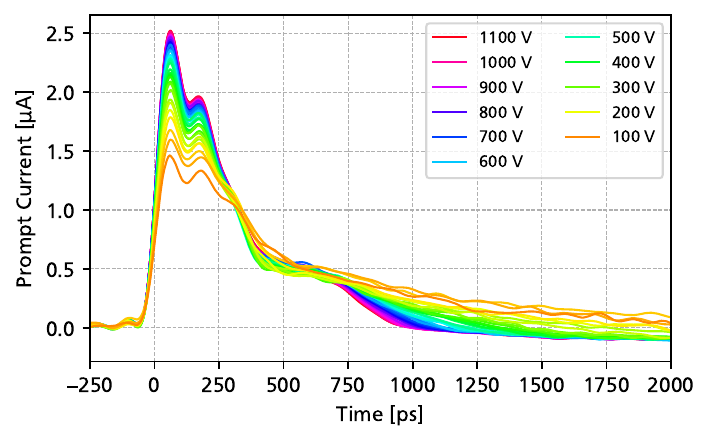}
    \caption{Averaged transient induced current for \SI{62.4}{\mega\electronvolt} proton beams as a function of the bias voltage. The signal amplitudes are scaled to the most probable value of the prompt current.}
    \label{fig:results:proton:waveforms}
\end{figure}

Figure~\ref{fig:results:proton:waveforms} shows the resulting averaged waveforms obtained using a \SI{62.4}{\mega\electronvolt} proton beam.
As expected, a good agreement with the data obtained using a UV-laser is observed.
A slight difference in the structure of the initial signal peak is observed, which can be attributed to the fact that protons deposit their charge very quickly ($< \SI{1}{\pico\second}$), which corresponds more closely to a Dirac-like excitation, while the pulse width of the UV-laser is of a similar time scale as the rise time of the readout electronics.

\subsection{Alpha Particles}
\begin{figure}[htp]
    \isPreprints{\centering}{} 
    \includegraphics[width=.625\textwidth]{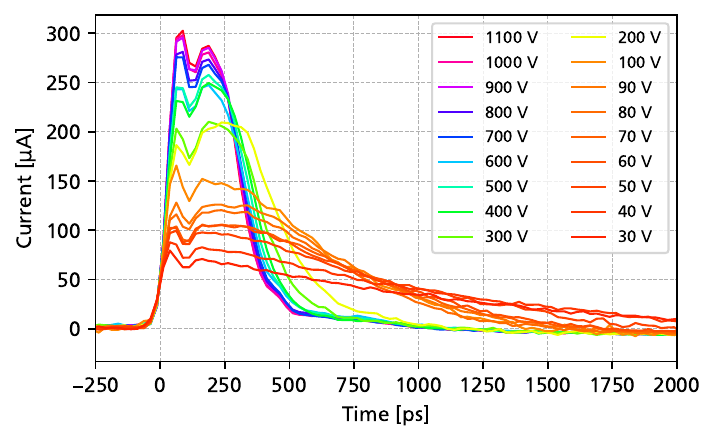}
    \caption{Transient currents for $^{241}\mathrm{Am}$ alpha particles as a function of the applied bias voltage.}
    \label{fig:results:alpha}
\end{figure}

Finally, measurements were performed using \SI{5.5}{\mega\electronvolt} alpha particles from a $^{241}\mathrm{Am}$ source in vacuum.
Alpha particles are readily absorbed in 4H-SiC, depositing their charge in the first \SI{15}{\micro\meter} of the detector.
Figure~\ref{fig:results:alpha} shows the measured transient currents as a function of the bias voltage.
For each voltage, only around \num{100} waveforms were averaged, as the acquisition rate was limited by the small active area of the detector and the low activity (\SI{1}{\kilo\becquerel}) of the radioactive source.
At high bias voltages, the waveforms are shorter than \SI{500}{\pico\second} (or \SI{300}{\pico\second} FWHM), as the slower holes only need to drift a short distance to the anode at the top of the device, compared to traversing the entire device thickness in the case of a uniform charge deposition.
This results in an almost rectangular current pulse, as expected by a selective injection of electrons, in clear contrast to the triangular pulses obtained for UV-TCT or proton beams.

\subsection{Saturation Velocities}
\label{sec:vsat_results}
\begin{figure}[htp]
    \isPreprints{\centering}{} 
    \begin{minipage}[b]{0.4\textwidth}
        \includegraphics[width=\textwidth]{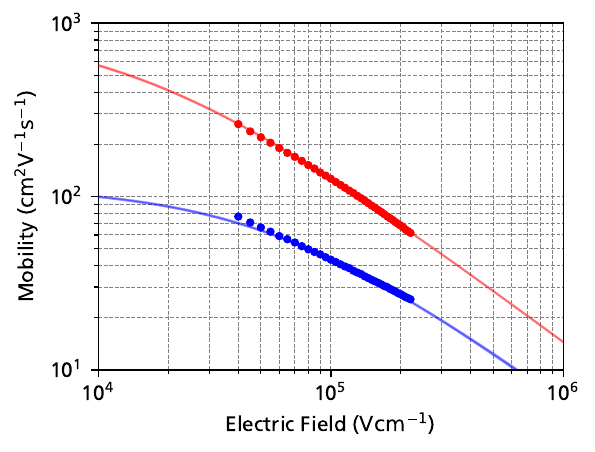}
        \quad 
        \centering (a)
    \end{minipage}
    \hspace{0.1\textwidth}
    \begin{minipage}[b]{0.4\textwidth}
        \includegraphics[width=\textwidth]{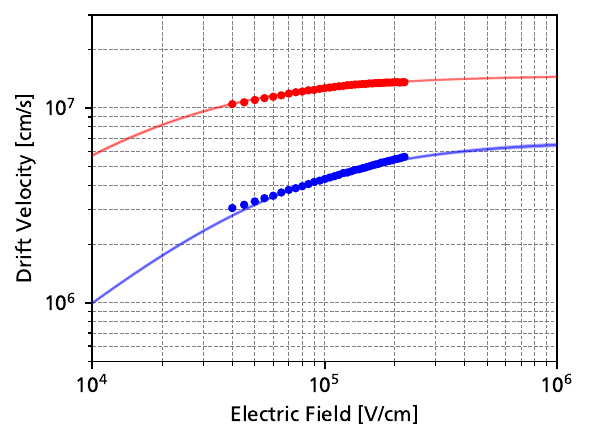}
        \quad 
        \centering (b)
    \end{minipage}
    \caption{Resulting mobility \textbf{(a)} and drift velocity \textbf{(b)} for electrons (red) and holes (blue).}
    \label{fig:results:mobility_fit}
\end{figure}

Figure~\ref{fig:results:mobility_fit} shows the drift velocity extracted from UV-TCT measurements as a function of the electric field together with a fit to the Caughey-Thomas high-field mobility model.
Saturation velocities of $\sim$\SI{1.47e7}{\centi\meter\per\second} for electrons and $\sim$\SI{0.69e7}{\centi\meter\per\second} for holes were obtained, with the full fit parameters shown in Table~\ref{tab:fit_results}.
These values are in accordance with values in the literature~\cite{burin2025}.
However, for the hole saturation velocity, almost all published values are based on Monte-Carlo simulations, and this measurement presents a valuable experimental data point for guiding TCAD simulations.
\begin{table}[htp]
    \isPreprints{\centering}{} 
    \caption{Fit results for electrons and holes using the Caughey-Thomas high-field mobility model~\eqref{eq:mobility}.}
    \begin{tabularx}{\textwidth}{CCC}
        \toprule
        \textbf{Parameter}           & \textbf{Electrons}                                                 & \textbf{Holes}                                                     \\
        \midrule
        $\mu_{\mathrm{low}}$ (fixed) & \SI{950}{\centi\meter^2\per\volt\second}~\cite{Ishikawa2023}       & \SI{115}{\centi\meter^2\per\volt\second}~\cite{Ishikawa2024}       \\
        $v_\mathrm{sat.}$            & \SI[separate-uncertainty=true]{1.47(2)e7}{\centi\meter\per\second} & \SI[separate-uncertainty=true]{0.69(4)e7}{\centi\meter\per\second} \\
        $\beta$                      & \num[separate-uncertainty=true]{0.96(1)}                           & \num[separate-uncertainty=true]{1.02(2)}                           \\
        \bottomrule
    \end{tabularx}
    \label{tab:fit_results}
\end{table}

\subsection{Comparison with Simulations}
\begin{figure}[htp]
    \isPreprints{\centering}{} 
    \begin{minipage}[b]{0.4\textwidth}
        \includegraphics[width=\textwidth]{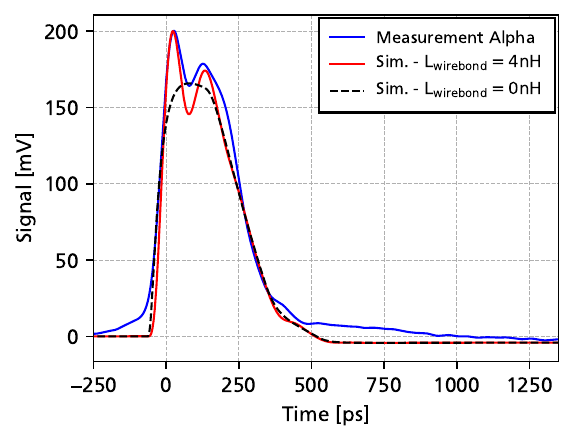}
        \quad 
        \centering (a)
    \end{minipage}
    \hspace{0.1\textwidth}
    \begin{minipage}[b]{0.4\textwidth}
        \includegraphics[width=\textwidth]{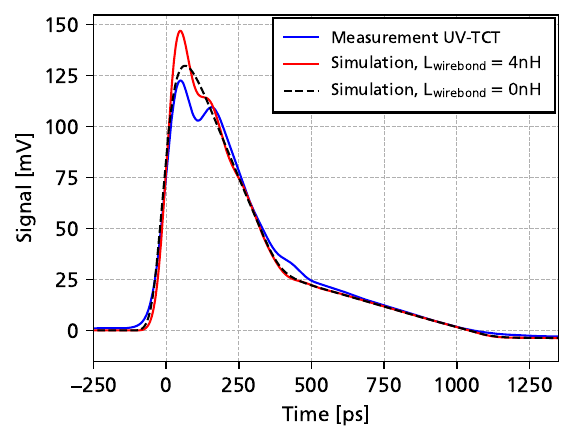}
        \quad 
        \centering (b)
    \end{minipage}
    \caption{Comparison of measurements and simulations, with and without the wire bond inductance, for alpha particles \textbf{(a)} and a pulsed UV-laser \textbf{(b)} at a bias voltage of \SI{500}{\volt}.}
    \label{fig:results:qucs}
\end{figure}

In Figure~\ref{fig:results:qucs}, a comparison of measurements and full simulations is shown.
For alpha particles, the simulation is able to predict the measured data quite well, while for the UV-TCT simulations, a slightly higher amplitude at the beginning of the signal is expected in the simulations.
These discrepancies affect mainly the high-frequency components, and evidently do not affect the slope of the waveform (used to fit charge carrier drift velocities) after around \SI{200}{\pico\second}.
Simulations are also shown for the case where the inductance of the wire bond is removed ($\mathrm{L_{wirebond} = \SI{0}{\nano\henry}}$), which removes the high-frequency $LC$-resonance oscillations.

\section{Discussion}
Although the nominal bandwidth of the LNA MMIC is \SI{10}{\giga\hertz}, the detector capacitance and parasitic impedances can reduce the bandwidth.
Together with the input impedance of the MMIC (\SI{50}{\ohm}), the detector capacitance forms an $RC$ low-pass filter.
However, with the detector capacitance of \SI{75}{\femto\farad}, the cut-off frequency of this filter is well above \SI{10}{\giga\hertz}.
Furthermore, together with the detector capacitance, the wire bond inductance of \SI{4}{\nano\henry} will form an $LC$-filter, with a pole at $1/(2 \pi \sqrt{LC}) \approx \SI{9.2}{\giga\hertz}$.
Damped oscillations at this frequency are visible in simulations and measurements, e.g., Figure~\ref{fig:results:qucs}.
Furthermore, the inductance of the bond wire results in a limited effective bandwidth of \SI{6}{\giga\hertz}, calculated from a \SI{10}{\percent}--\SI{90}{\percent} rise time of \SI{58}{\pico\second}.

\begin{figure}[htp]
    \isPreprints{\centering}{} 
    \begin{minipage}[b]{0.4\textwidth}
        \includegraphics[width=\textwidth]{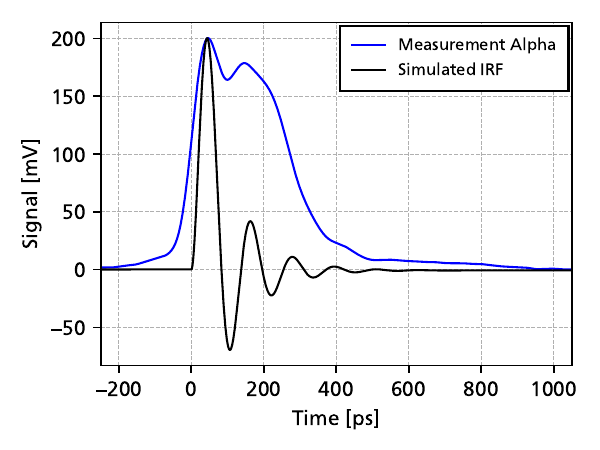}
        \quad 
        \centering (a)
    \end{minipage}
    \hspace{0.1\textwidth}
    \begin{minipage}[b]{0.4\textwidth}
        \includegraphics[width=\textwidth]{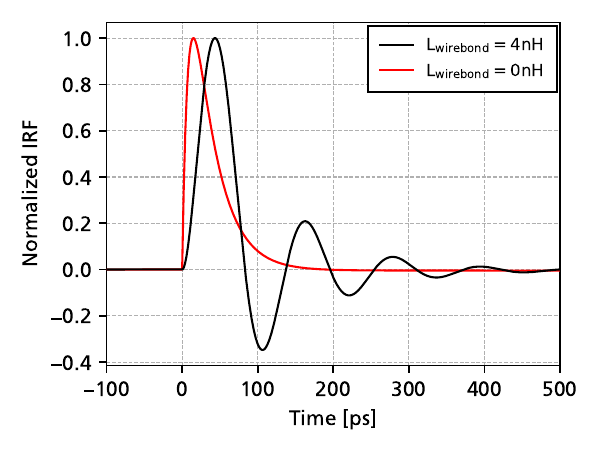}
        \quad 
        \centering (b)
    \end{minipage}
    \caption{\textbf{(a)}: Alpha particle measurements shown together with the simulated impulse response function (IRF) of the readout board. \textbf{(b)}: Comparison of the IRF with and without the effect of the bond wire inductance.}
    \label{fig:results:qucs_IRF}
\end{figure}
Figure~\ref{fig:results:qucs_IRF}(a) shows the simulated impulse response function (IRF) of the full circuit model, obtained using a Dirac-like input pulse.
In comparison to the transient current of the detector (shown for the example of alpha particles), the IRF is much shorter in time, which implies that the readout electronics are sufficiently fast to resolve the transient current of the detector.
Figure~\ref{fig:results:qucs_IRF}(b) shows a comparison of the full  IRF with the case where the wire bond inductance is removed.
Not only does the wire bond introduce significant oscillations, but it also limits the readout bandwidth, as evidenced by an increased rise time.

\begin{figure}[htp]
    \isPreprints{\centering}{} 
    \includegraphics[width=.625\textwidth]{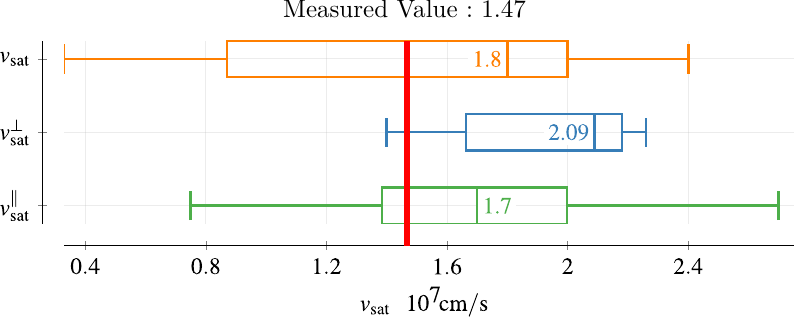}
    \caption{Comparison of the measured electron saturation drift velocity with literature values~\cite{burin2025}. The literature values are given for an unspecified crystal direction ($v_{\mathrm{sat}}$), perpendicular to the $c$-axis ($v_{\mathrm{\perp}}$), and parallel to the $c$-axis ($v_{\mathrm{\parallel}}$).}
    \label{fig:discussion:literature}
\end{figure}

The measured electron saturation velocity of \SI[separate-uncertainty=true]{1.47(2)e7}{\centi\meter\per\second} compares well with literature values~\cite{burin2025}, as depicted in Figure~\ref{fig:discussion:literature}.
For the hole saturation velocity, only a few reports exist in the literature, with the majority of these being Monte-Carlo simulations.
However, the measured hole saturation velocity of \SI[separate-uncertainty=true]{0.69(4)e7}{\centi\meter\per\second} lies inside of the range of reported values, from \SI{0.65e7}{\centi\meter\per\second} to \SI{0.86e7}{\centi\meter\per\second}~\cite{Nilsson2000, Zhao2000, Hjelm2003}.

In the UV-TCT laser measurements, charge was injected between the pad and the charge collection (guard) ring.
As this ring was left floating, the electric potential is expected to be identical to the pad (contact of the diode), therefore resulting in a uniform electric field.
As a result, these measurements are expected to yield accurate results for the charge carrier drift velocities, even though the charge is not injected directly in the center of the device.
As a cross-check, the analysis of charge carrier drift velocities based on the transient current waveforms was also performed for data obtained using a beam of \SI{62.4}{\mega\electronvolt} protons.
Here, saturation velocities of \SI[separate-uncertainty=true]{1.54(2)e7}{\centi\meter\per\second} for electrons and \SI[separate-uncertainty=true]{0.55(4)e7}{\centi\meter\per\second} for holes were obtained, with $\beta_e = \num[separate-uncertainty=true]{1.01(2)}$ and $\beta_h=\num[separate-uncertainty=true]{1.36(3)}$.
If $\beta$ is fixed to the values obtained from UV-TCT measurements (see Table~\ref{tab:fit_results}), the obtained saturation velocities are \SI[separate-uncertainty=true]{1.58(2)e7}{\centi\meter\per\second} for electrons and \SI[separate-uncertainty=true]{0.68(4)e7}{\centi\meter\per\second} for holes.
Both approaches are comparable to the UV-TCT measurement, albeit with slightly higher drift velocities for electrons and slightly lower for holes.
It is also possible that the error bars of the drift velocities extracted using the proton data are larger, because the transient current waveform in the proton beam measurements is estimated by averaging a small signal subset.

\section{Conclusion}
A high-frequency readout capable of resolving the transient current in \SI{50}{\micro\meter} thin 4H-SiC PiN diodes was developed and applied to measurements using a pulsed UV laser, alpha particles, and a \SI{62.4}{\mega\electronvolt} proton beam.
The high effective readout bandwidth of \SI{6}{\giga\hertz} was used to measure the electron and hole drift velocities and to fit a high-field mobility model.
From pulsed UV laser measurements, saturation drift velocities parallel to the $c$-axis of \SI[separate-uncertainty=true]{1.47(2)e7}{\centi\meter\per\second} for electrons and \SI[separate-uncertainty=true]{0.69(4)e7}{\centi\meter\per\second} for holes were obtained, providing valuable input to TCAD simulations.
In order to understand the readout electronics in detail and to investigate the effect of parasitic impedances such as the wire bond inductance, simulations leveraging TCAD, Allpix$^2$, and QUCS were performed, which agree well with the measurements.
Future studies will aim at applying the developed readout electronics to other detector types, such as silicon-based detectors and low-gain avalanche diodes (LGADs), and will exploit more targeted charge injection methods, such as two-photon absorption.

\vspace{6pt}
\authorcontributions{Conceptualization, A.G., S.O., S.G.; Data curation, A.G., S.O., S.G.; Formal analysis, A.G., S.O., S.G.; Funding acquisition, T.B.; Investigation, A.G., S.O., S.G.; Methodology, A.G., S.O, S.G.; Project administration, T.B.; Resources, T.B.; Software, A.G., S.O.; Supervision, S.G., T.B.; Validation, A.G., S.O., J.B., S.G., M.K., D.R., S.W., T.B.; Visualization, A.G, S.O, S.G.; Writing - original draft, A.G., S.O., S.G.; Writing - review \& editing, A.G., S.O., J.B., S.G., M.K., D.R., S.W., T.B. }

\funding{
    The authors thank the Institute of Microelectronics of Barcelona, IMB-CNM-CSIC, for supporting the production of the 4H-SiC detectors.
    The financial support of the Austrian Ministry of Education, Science and Research is gratefully acknowledged for providing beam time and research infrastructure at MedAustron.
    This work was supported by the Austrian Research Promotion Agency (FFG) in the project "RadHardDetSim" (895291).
    The authors acknowledge TU Wien Bibliothek for financial support through its Open Access Funding Programme.
}

\dataavailability{The raw data supporting the conclusions of this article will be made available by the authors on request.}

\acknowledgments{We want to acknowledge the DRD3 R\&D collaboration, especially working group 6, for continuous discussions and feedback. We also thank Andreas Bauer for bonding the detector devices and Klaus-Dieter Fischer for PCB assembly work.}

\conflictsofinterest{The authors declare no conflicts of interest.}

\appendixtitles{no}

\isPreprints{}{
    \begin{adjustwidth}{-\extralength}{0cm}
        } 

        \reftitle{References}
        \bibliography{bibliography.bib}
        \isPreprints{}{
    \end{adjustwidth}
} 
\end{document}